 \documentstyle[12pt]{article}
 
\def\appendix#1{
  \addtocounter{section}{1}
  \setcounter{equation}{0}
  \renewcommand{\thesection}{\Alph{section}}
  \section*{Appendix \thesection\protect\indent \parbox[t]{11.715cm} {#1} }
  \addcontentsline{toc}{section}{Appendix \thesection\ \ \ #1}
  }

\newcommand{\newsection}{
\setcounter{equation}{0}
\section}
\def\tr{\,{\rm tr}\,}
\def\Tr{\,{\rm Tr}\,}
\def\Sp{\,{\rm Sp}\,}
\def\E{{\,\rm e}\,}
\newcommand{\rf}[1]{(\ref{#1})}
\newcommand{\orr}[1]{\overrightarrow{#1}}
\newcommand{\ol}[1]{\overleftarrow{#1}}
\newcommand{\vc}[1]{\mbox{\boldmath$#1$}}
\def\b{\beta}
\def\m{\mu}
\def\x{\xi}
\def\mod{{\rm mod}}
\def\sign{\,{\rm sign}}
\def\s{\sigma}
\def\d{\partial}
\def\D{\delta}
\def\L{\Delta}
\def\O{{\cal O}}

\def\G{{\cal G}}
\def\J{{\cal J}}
\def\e{\varepsilon}
\def\g{\Gamma}
\def\a{\eta}
\def\dq{\triangle\omega}
\def\o{\omega}
\def\l{\lambda}
\def\r{\Lambda}
\def\z{\zeta^{\prime}(-2)}
\def\dl{\overrightarrow{D}_{\m}}
\def\dr{\overleftarrow{D}_{\m}}
\def\<<{\stackrel{<}{\sim}}
\begin{document}

\begin{titlepage}
\begin{flushright}
SMI-95-4 \\
October, 1995 \\
hep-th/9510031
\end{flushright}
\vspace{1.25cm}

\begin{center}
{\LARGE Monopole determinant in Yang--Mills theory}
\end{center}
\vspace{0.2cm}
\begin{center}
{\LARGE at finite temperature}
\end{center}
\vspace{1cm}
\begin{center}
{\large K.\ Zarembo}
\footnote{E--mail:   \ zarembo@class.mian.su \ }\\
 \mbox{} \\ {\it Steklov Mathematical Institute,} \\
{\it Vavilov st. 42, GSP-1, 117966 Moscow, RF}
\end{center}

\vskip 1 cm
\begin{abstract}
The fluctuation determinant in the BPS monopole background is calculated
 in the finite--temperature $SU(2)$ gauge theory.
 \end{abstract}

\vspace{1cm}
\noindent

\vspace*{\fill}\pagebreak
\end{titlepage}

\section{Introduction}

 It is well known that asymptotically free gauge theories become weakly
 coupled at high temperature because of the decrease of the running
 coupling constant $g^2(T)$ with $T$. But due to infrared
 divergences the domain of applicability of perturbation theory is limited
 by the few lowest orders and by distances much smaller than
 $\left(g^2T\right)^{-1}$, beyond which nonperturbative effects become
 important. On the other hand, the smallness of the effective coupling
 justify an application of semiclassical methods, which provide
 essentially nonperturbative information. While the finite temperature
 instanton calculations have been performed in detail \cite{GPY}, the
 contribution of another type of finite action classical solutions, the
 BPS monopoles \cite{BPS}, was studied only numerically \cite{SV}.
 Analogously
 to the 't~Hooft--Polyakov monopoles in three dimensional adjoint Higgs
 model, which generate a mass gap and an area law for Wilson loops
 \cite{Polyakov}, the BPS monopoles contribute, in principle, to the
 spatial string tension and to the magnetic screening, responsible for the
 infrared divergences of perturbation theory.

 The mean density of the BPS monopoles is determined in a dilute gas
 approximation by the contribution of a single monopole to the partition
 function. The calculation of the latter implies an account of the
 fluctuations in the monopole background. The purpose of the present paper
 is to calculate the corresponding fluctuation determinant. The method we
 use is based on the fact that the propagators in the monopole background
 are known explicitly \cite{Adler}--\cite{Nahm80}. This method allows to
 find completely the functional dependence of the determinant on the
 monopole size. However, the general expressions are too cumbersome, and
 only the results for the monopoles of a small size will be exhibited
 explicitly.

 It is not evident that the semiclassical calculations for the BPS
 monopoles can at all be performed. The point is that the Coulomb--like
 behavior of the monopole field may lead to the logarithmic infrared
 divergences at a one--loop level. Such divergences indeed have been found
 in a similar problem of calculation of the quantum corrections to the
 mass of the 't~Hooft--Polyakov monopole \cite{KS}. The observation that
 the BPS monopole can be obtained as an infinite size limit of the
 periodic instanton \cite{Rossi,GPY} does not help, since the instanton
 determinant is singular in this limit. However, this singularity is a
 consequence of the fact that the leading contribution to the determinant
 comes from the distances much larger than the instanton size \cite{GPY},
 where the field of the instanton differs considerably from that of the
 monopole. The monopole determinant, as it will be shown, is finite and
 free from infrared divergences.

\newsection{Monopole contribution to partition function\label{mnpl}}

In a standard formulation of a field theory at finite temperature the
partition function and correlators are represented in a form of Euclidean
functional integrals with periodic boundary conditions in an imaginary
time with a period, equal to the inverse temperature $\b=1/T$. The static
fields are obviously periodic and thus contribute to the functional
integral. We consider pure Yang--Mills theory with gauge group $SU(2)$. It
possesses (anti)self--dual static classical solutions of a monopole type
\cite{BPS}:
\[
A_0^a=q\m n_a\left(\coth\m r-\frac{1}{\m r}\right),
\]
\[
A_i^a=\e_{aij}n_j\frac{1}{r}\left(1-\frac{\m r}{\sinh\m r}\right),
\]
\[
F_{i0}^a=qn_an_i\frac{1}{r^2}\left[1-\frac{(\m r)^2}{\sinh^2\m r}\right]
+q\m^2\left(\D_{ai}-n_an_i\right)\frac{1}{\sinh\m r}
\left(\coth\m r-\frac{1}{\m r}\right),
\]
\begin{equation}\label{monopole}
F_{ij}^a=-q\e_{ijk}F_{k0}^a,
\end{equation}
where $n_i=x_i/r$, $r=|{\bf x}|$ and $q=\pm 1$ is a magnetic charge of the
solution; in what follows the monpoles with $q=1$ are considered.
We shall also use a
matrix notation for gauge potentials:  $A_{\m}=A_{\m}^a\s^a/2i$.

It is worth mentioning that $A_0$ in \rf{monopole} does not decrease at
infinity. Consideration of a constant background shows \cite{GPY} that
not all such fields, but only that, satisfying the condition
\begin{equation}\label{pexp}
\lim_{|{\bf x}|\rightarrow\infty}\,{\rm P}\exp\int_0^{\b}
dx_0\,A_0(x_0,{\bf x})=\pm{\bf 1},
\end{equation}
give finite contribution to the partition function in the thermodynamic
limit, others being suppressed by a factor $\exp(-cVT^3)$. The condition
\rf{pexp} leads to a quantization of the monopole size:
\begin{equation}\label{mk}
\m=\frac{2\pi k}{\b}.
\end{equation}
The integer $k$ is nothing but the topological charge of a monopole:
\[
k=\frac{q}{32\pi^2}\int_0^{\b}dx_0\,\int d^3x\,\frac12\e_{\m\nu\l\rho}
F_{\m\nu}^aF_{\l\rho}^a.
\]
Only the solutions with $\m$, given by \rf{mk},  can be gauge transformed
to the periodic fields decreasing at infinity. This gauge transformation
is periodic for even and antiperiodic for odd $k$. When the matter fields
are added, only the periodic gauge transformations are allowed, so in this
case the topological charge of a monopole should be even. The possibility
of antiperiodic gauge transformations in pure gauge theory is a
consequence of ${\bf Z}_2$ symmetry \cite{z2}. However, at high
temperature ${\bf Z}_2$ symmetry is spontaneously broken, thus the
contribution of the monopoles with odd topological charge apparently
should not be taken into account in this case as well.

The monopole density can be calculated by standard semiclassical
techniques. The zero modes are treated by collective coordinate procedure.
Since the BPS monopole is self--dual, the contribution of non--zero modes
can be expressed through the spin-$0$ fluctuation determinant \cite{BC}.
In a background gauge $D_{\m}^{({\rm cl})}A_{\m}=0$ it gives, together
with a ghost contribution, a factor
$\left[\det\left(-D^2\right)\right]^{-1}$, where the covariant derivative
acts in the adjoint representation:
$D_{\m}^{ab}=\D^{ab}\d_{\m}+\e^{acb}A_{\m}^{c}$. Thus one obtains for a
 one--monopole partition function:
\begin{equation}\label{z1} Z_1=\sum_k\int
d\nu\,J^{1/2}\left[\det\left(-D^2\right)\right]^{-1} \E^{-8\pi^2k/g_0^2},
\end{equation}
where the integration is over collective coordinates,
corresponding to the zero modes. The Jacobian $J$ ensures a correct
normalization of the measure and depends on a parametrization of the
 classical solutions.
The last factor is a
contribution of the  classical action.

There are four evident zero modes in the monopole background. Three of
them correspond to the global translations and one -- to the global gauge
 transformation, commuting with $A_0$. Let us stress that there is no zero
 mode, associated with dilatations -- the integration over $\m$ is
 replaced by summation over $k$. However, the total number of the zero
 modes is larger, it grows linearly with $k$, so the BPS monopole
 \rf{monopole} is a particular case of a more general periodic solution
 with a unit magnetic charge and topological charge $k$. Such solutions
 are generally nonstatic and, in principle, all of them can be classified
 \cite{Nahm83}. More detailed treatment of the zero mode contribution is
 beyond the scope of the present paper.

 The monopole density is ultraviolet divergent due to the contribution
 of both fluctuations and zero modes. After a proper regularization all
 ultraviolet infinities are absorbed by a standard one--loop coupling
 constant renormalization, and the divergent contribution reduces to the
 replacement of the bare coupling $g_0^2$ in \rf{z1} by the renormalized
 one $g^2(T)$.

 For the monopoles with nonintegral topological charge the fluctuation
 determinant contains also the infrared
 divergences. We regularize them by cutting divergent integrals on
 an upper bound. The leading divergence is proportional to the volume, it
 produce the factor $\exp(-cVT^3)$, which leads to the abovementioned
 quantization of a monopole size. This divergence, as well as the
 nonleading power--like and logarithmic ones, cancel, when the condition
  \rf{mk} is satisfied.

 We shall perform explicit calculations of the determinant for large $k$,
 systematically dropping $O(1/k)$ terms. Here we quote the final result:
 \begin{equation}\label{detd2}
 -\ln\det\left(-D^2\right)=\frac23k\ln \frac{T}{\r}+\frac23k\ln
 k-0.240373k+O\left(k^0\right),
 \end{equation}
 where $\r$ is ultraviolet cutoff.

\newsection{Variation of determinant}

 We use the following representation for the regularized determinant:
 \begin{equation}\label{rdet}
 \g\equiv -\ln\det\left(-D^2\right)=\int_{1/\r^2}^{\infty}\frac{dt}{t}\,
 \Sp\E^{tD^2}.
 \end{equation}
 The variation of $\g$ can be expressed through the Green function of
 $\left(-D^2\right)$. We choose $\m$ to be a parameter of variation, then
 \begin{equation}\label{varf}
 \m\frac{\d A_{\nu}}{\d\m}=A_{\nu}+x_j\d_jA_{\nu}=
 x_jF_{j\nu}+D_{\nu}(x_jA_j)+\D_{\nu 0}A_0=x_jF_{j\nu}+\D_{\nu 0}A_0
 \end{equation}
 and
 \begin{equation}\label{vargamma}
 \m\frac{\d\g}{\d\m}=\frac12\int_{1/\r^2}^{\infty}\frac{dt}{t}\,
 \int_0^t ds\,\Sp\left(\E^{sD^2}\m\frac{\d D^2}{\d\m}\E^{(t-s)D^2}
 +\E^{(t-s)D^2}\m\frac{\d D^2}{\d\m}\E^{sD^2}\right).
 \end{equation}
 Consider the integrand in this expression:
 \begin{equation}\label{defcf}
 {\cal F}(t,s)=\Sp\left(\E^{sD^2}\m\frac{\d D^2}{\d\m}\E^{(t-s)D^2}
 +\E^{(t-s)D^2}\m\frac{\d D^2}{\d\m}\E^{sD^2}\right).
 \end{equation}
 At first sight it does not depend on $s$ due to the cyclic property of
 the trace, equivalent for differential operators to integration by parts.
 This, however, leaves boundary terms and, as a result, ${\cal F}$ is a
 second order polynomial in $s$. To show it, let us differentiate
 \rf{defcf} $n$ times with respect to $s$:
 \begin{equation}\label{der}
 \frac{\d^n}{\d s^n}{\cal F}(t,0)=\Sp\left({\cal O}_{n+1}
 \E^{tD^2}+(-1)^n\E^{tD^2}{\cal O}_{n+1}\right).
 \end{equation}
 Here ${\cal O}_n$ is a differential operator of $n$--th order, defined
 recursively by
 \begin{equation}\label{o1}
 {\cal O}_1=\m\frac{\d D^2}{\d\m}=\left\{\m\frac{\d A_{\nu}}{\d\m}
 ,D_{\nu}\right\},
 \end{equation}
 \begin{equation}\label{on}
 {\cal O}_{n+1}=\left[D^2,{\cal O}_n\right].
 \end{equation}
 It is not difficult to write down explicitly the first few ones:
 \begin{equation}\label{o2}
 \O_2=\left\{D_{\nu},\left\{\m\frac{\d
 A_{\l}}{\d\m},F_{\nu\l}\right\}\right\}
 +\left\{D_{\nu},\left\{\left(D_{\nu}\m\frac{\d
 A_{\l}}{\d\m}\right),D_{\l}\right\}\right\},
 \end{equation}
 \begin{eqnarray}\label{o3}
 \O_3&=&\left\{D_{\l},
 \left\{D_{\nu}  ,\left\{\m\frac{\d A_{\rho}}{\d\m}
 ,\left(D_{\l}F_{\nu\rho}\right) \right\}\right\}
 +
 \left\{F_{\l\nu}  ,\left\{
 \m\frac{\d A_{\rho}}{\d\m}  ,F_{\nu\rho}  \right\}\right\}
 \right.\nonumber\\
 &&+
 \left\{D_{\nu}  ,\left\{
 \left(D_{\l}\m\frac{\d A_{\rho}}{\d\m}\right)  ,
 F_{\nu\rho}  \right\}\right\}
 +\left\{F_{\l\nu}  ,\left\{
 \left(D_{\nu}\m\frac{\d A_{\rho}}{\d\m}\right)  ,D_{\rho}
 \right\}\right\}
 \nonumber\\&&\left.
 +\left\{D_{\nu}  ,\left\{
 \left(D_{\l}D_{\nu}\m\frac{\d A_{\rho}}{\d\m}\right)  ,
 D_{\rho}  \right\}\right\}
 +\left\{D_{\nu}  ,\left\{
 \left(D_{\nu}\m\frac{\d A_{\rho}}{\d\m}\right)  ,
 F_{\l\rho}  \right\}\right\}
 \right\}.
 \end{eqnarray}

 It follows from the definition of $\O_{n+1}$ that for $n\geq 1$ the right
 hand side of eq. \rf{der} is given by the trace of a commutator or,
 equivalently, by an integral of a total derivative, which reduces to a
 surface term:
 \begin{eqnarray}\label{sfint}
 \frac{\d^n}{\d s^n}{\cal F}(t,0)&=&\Sp\left[D^2,{\cal O}_{n}
 \E^{tD^2}+(-1)^n\E^{tD^2}{\cal O}_{n}\right]
 \nonumber\\
 &=&\int_{0}^{\b}dx_0\,\int d^3x\,\Tr\left[
 \orr{D}^2
 \left(\orr{\O}_n{\cal K}(x,y;t)+(-1)^n{\cal K}(x,y;t)\ol{\O}_n\right)
 \right.\nonumber\\
 &&
 \left.\left.
 -\left(\orr{\O}_n{\cal K}(x,y;t)+(-1)^n{\cal K}(x,y;t)\ol{\O}_n\right)
 \ol{D}^2\right]\right|_{y=x}
 \nonumber\\
 &=&\int_{0}^{\b}dx_0\,\oint_{S_R} d^2\Sigma\,n_i\Tr\left[
 \orr{D}_i
 \left(\orr{\O}_n{\cal K}(x,y;t)+(-1)^n{\cal K}(x,y;t)\ol{\O}_n\right)
 \right.\nonumber\\
 &&
 \left.\left.
 +\left(\orr{\O}_n{\cal K}(x,y;t)+(-1)^n{\cal K}(x,y;t)\ol{\O}_n\right)
 \ol{D}_i\right]\right|_{y=x},
 \end{eqnarray}
 where $\Tr$ denotes the trace over color indices in the adjoint
 representation and $R\rightarrow\infty$.
 By ${\cal K}$ we denote a heat kernel of
 $-D^2$:
 \begin{equation}\label{heatk}
 {\cal K}(x,y;t)=\left\langle x\left|\E^{tD^2}\right|y\right\rangle.
 \end{equation}
 By noting that $D_{\nu}\m\frac{\d A_{\l}}{\d\m}$ falls exponentially at
 infinity and $F_{\nu\l}$ is of order $1/r^2$ we conclude that the
 coefficients of the operator $\O_3$, and thus of all $\O_n$ with $n\geq
 3$, are, at least, of order $1/r^3$. As a consequence, the surface
 integral \rf{sfint} vanishes for $n\geq 3$ and only the first and the
 second derivatives of ${\cal F}$ are not equal to zero.

 Using an explicit expressions for $\O_1$ and $\O_2$ and taking into
 account that only the variation of $A_0$ does not decrease at infinity we
 find:
 \begin{eqnarray}\label{der1}
 \frac{\d}{\d s}{\cal F}(t,0)&=&2
 \int_{0}^{\b}dx_0\,\oint_{S_R} d^2\Sigma\,n_i\Tr\left[
 \m\frac{\d A_0}{\d\m}
 \left(\orr{D}_i\orr{D}_0{\cal K}(x,y;t)
 \right.\right.\nonumber\\&&\left.\left.\left.
 -\orr{D}_i{\cal K}(x,y;t)\ol{D}_0
 +\orr{D}_0{\cal K}(x,y;t)\ol{D}_i
 -{\cal K}(x,y;t)\ol{D}_0\ol{D}_i
 \right)\right|_{y=x}\right] \nonumber\\
 &=&4\int_{0}^{\b}dx_0\,\oint_{S_R} d^2\Sigma\,n_i\Tr\left(
 \m\frac{\d A_0}{\d\m}F_{i0}
 {\cal K}(x,x;t)\right),
 \end{eqnarray}
 \begin{eqnarray}\label{der2}
 \frac{\d^2}{\d s^2}{\cal F}(t,0)&=&4
 \int_{0}^{\b}dx_0\,\oint_{S_R} d^2\Sigma\,n_i\Tr\left[
 \m\frac{\d A_0}{\d\m}F_{j0}
 \left(\orr{D}_i\orr{D}_j{\cal K}(x,y;t)
 \right.\right.\nonumber\\&&\left.\left.\left.
 +\orr{D}_i{\cal K}(x,y;t)\ol{D}_j
 +\orr{D}_j{\cal K}(x,y;t)\ol{D}_i
 +{\cal K}(x,y;t)\ol{D}_j\ol{D}_i
 \right)\right|_{y=x}\right] \nonumber\\
 &=&-\frac{8}{t}\int_{0}^{\b}dx_0\,\oint_{S_R} d^2\Sigma\,n_i\Tr\left(
 \m\frac{\d A_0}{\d\m}F_{i0}
 {\cal K}(x,x;t)\right).
 \end{eqnarray}
 The last equality holds, because at infinity
 \begin{equation}\label{heatd}
 \orr{D}_i\orr{D}_j{\cal K}(x,y;t)=\left[-\frac{1}{2t}\delta_{ij}
 +\frac{1}{4t^2}(x-y)_i(x-y)_j\right]{\cal K}(x,y;t)
 +O\left(\frac{1}{r}\right)
 \end{equation}
 and analogous formulas are valid for
 $\orr{D}_i{\cal K}(x,y;t)\ol{D}_j$ and
 ${\cal K}(x,y;t)\ol{D}_j\ol{D}_i$. They can be derived directly from the
 defining equation for the heat kernel with the use of it's asymptotic
 form at $t\rightarrow 0$:
 \begin{equation}\label{kt0}
 {\cal K}^{ab}(x,y;t)\sim\frac{\delta^{ab}}{16\pi^2t^2}
 \E^{-(x-y)^2/4t}.
 \end{equation}

 Thus we obtain:
 \begin{eqnarray}\label{cf}
 {\cal F}(t,s)&=&2\Sp\left(\m\frac{\d D^2}{\d\m}\E^{tD^2}\right)
 \nonumber\\&&+4\left(s-\frac{s^2}{t}\right)\int_{0}^{\b}dx_0\,
 \oint_{S_R} d^2\Sigma\,n_i\Tr\left(
 \m\frac{\d A_0}{\d\m}F_{i0}
 {\cal K}(x,x;t)\right).
 \end{eqnarray}
 Substituting this expression in the variation of the determinant, eq.
 \rf{vargamma}, and doing the integral over $s$ one finds:
 \begin{eqnarray}\label{mde}
 \m\frac{\d\g}{\d\m}&=&\int_{1/\r^2}^{\infty}dt\,
 \Sp\left(\m\frac{\d D^2}{\d\m}\E^{tD^2}\right)
 \nonumber\\&&+\frac13\int_{1/\r^2}^{\infty}dt\,t
 \int_{0}^{\b}dx_0\,\oint_{S_R} d^2\Sigma\,n_i\Tr\left(
 \m\frac{\d A_0}{\d\m}F_{i0}
 {\cal K}(x,x;t)\right).
 \end{eqnarray}
 After the change of variables $t\rightarrow t-1/\r^2$ the integration
 over $t$ can be easily performed. As it was shown in ref. \cite{CGOT},
 the first term in \rf{mde} remains finite in $\r\rightarrow\infty$ limit
 and is given by
 \begin{equation}\label{vardet}
 \m\frac{\d\g_I}{\d\m}
 =\int_0^{\b}dx_0\,
 \int d^3x\,\e^{acb}
 \m\frac{\d A_{\nu}^c}{\d\m}\J^{ba}_{\nu},
 \end{equation}
 \begin{equation}\label{defj}
 \J_{\nu}^{ba}(x)=\left.\left(\orr{D}_{\nu}^{bd}\G_R^{da}(x,y)
 +\G_R^{bd}(y,x)
 \ol{D}_{\nu}^{da}\right)
 \right|_{y=x},
 \end{equation}
 where $\G_R$ is the Green function, $\G=\left(-D^2\right)^{-1}$,
 regularized by the point splitting:
 \begin{equation}\label{grdef}
 \G_{R}^{ab}(x,y)=\G^{ab}(x,y)
 -\frac{\frac12\tr\s^a\Phi(x,y)\s^b\Phi(y,x)}{4\pi^2(x-y)^2},
 \end{equation}
 \begin{equation}\label{pexpd}
 \Phi(x,y)={\rm P}\exp\int_x^y dx'_{\m}\,A_{\m}(x').
 \end{equation}
 The second term in \rf{mde} reduces to a sum of two surface integrals:
 \begin{equation}\label{ga2}
 \m\frac{\d\g_{II}}{\d\m}=
 \frac13\int_{0}^{\b}dx_0\,\oint_{S_R} d^2\Sigma\,n_i\Tr\left[
 \m\frac{\d A_0}{\d\m}F_{i0}
 \left\langle x\left|\left(-D^2\right)^{-2}
 \E^{D^2/\r^2}\right|x\right\rangle\right]
 \end{equation}
 and
 \begin{equation}\label{ga3}
 \m\frac{\d\g_{III}}{\d\m}=
 \frac13\int_{0}^{\b}dx_0\,\oint_{S_R} d^2\Sigma\,n_i\Tr\left[
 \m\frac{\d A_0}{\d\m}F_{i0}
 \frac{1}{\r^2}\left\langle x\left|\left(-D^2\right)^{-1}
 \E^{D^2/\r^2}\right|x\right\rangle\right].
 \end{equation}
 In the last expression $\left(-D^2\right)^{-1}$ can be replaced by a free
 propagator and the asymptotic form \rf{kt0} of the heat kernel can be
 used. One gets in the $\r\rightarrow\infty$ limit
 \[\frac{1}{\r^2}\left\langle x\left|\left(-D^2\right)^{-1}
 \E^{D^2/\r^2}\right|x\right\rangle
 \sim\frac{1}{\r^2}\int
 d^4y\,\frac{1}{4\pi^2(x-y)^2}\frac{\r^4}{16\pi^2}\E^{-(y-x)^2\r^2/4}
 =\frac{1}{16\pi^2},\]
 and after simple transformations, involving the use of the equations of
 motion, \rf{ga3} can be rewritten as
 \[
 \m\frac{\d\g_{III}}{\d\m}=
 \frac13\frac{1}{64\pi^2}\m\frac{\d}{\d\m}\int_{0}^{\b}dx_0\,
 \int d^3x\,\Tr F_{\nu\l}F_{\nu\l}=-\frac13\m\frac{\d k}{\d\m}.\]
 Thus
 \begin{equation}\label{ga3main}
 \g_{III}=-\frac{1}{3}k.
 \end{equation}

 Eqs. \rf{vardet}, \rf{defj} and \rf{ga2} express the variation of
 $\g_{I}$ and $\g_{II}$ in terms of the isospin--$1$ propagator in the
 monopole background. It is convenient to use the ADHM construction for
 self--dual gauge fields \cite{ADHM}, developed for the BPS monopoles in
 ref. \cite{Nahm80}; which provides the expressions for propagators in a
 rather compact form \cite{CFGT,CWS}. It is worth mentioning that the ADHM
 construction enables to calculate the multi--instanton determinants in
 complete generality \cite{CGOT,det}.

\newsection{Ultraviolet finite part of determinant\label{sec}}

 The essence of the ADHM construction is introduction of an auxiliary
 linear space; that of $2\times 2$ matrix--valued functions of a new
 variable $z\in [-1/2,1/2]$ in the case of $SU(2)$ monopoles. The scalar
 product in this space is defined by \begin{equation}\label{scprod}
 \langle v_1|v_2\rangle=
 \int_{-1/2}^{1/2}dz\,v_1^{\dagger}(z)v_2(z),
 \end{equation}
 where matrix multiplication is implied. Thus $\langle v_1|
 v_2\rangle$
 is a $2\times 2$ matrix. The one--monopole solution is given in these
 terms by
 \begin{equation}\label{ADHM}
 A_{\m}(x)=\langle v(x)|\d_{\m}v(x)\rangle,
 \end{equation}
 where \cite{Nahm80}
 \begin{equation}\label{v}
 \left|v(x)\right\rangle=\left(\frac{\x}{\sinh\x}\right)^{1/2}
 \exp\left(i\x_0z-\x_i\s^iz\right)
 \end{equation}
 is a solution to the equation
 \begin{equation}\label{Dv}
 \L^{\dagger}(x)\left|v(x)\right\rangle\equiv
 \left(i\frac{d}{dz}+\x_0+i\x_i\s^i\right)\left|v(x)\right\rangle=0,
 \end{equation}
 normalized by a condition
 \begin{equation}\label{norm}
 \langle v(x)|v(x)\rangle={\bf 1}.
 \end{equation}
 Here and in what follows we denote by $\x_{\m}$ the rescaled coordinate:
 \begin{equation}\label{x}
 \x_{\nu}=\m x_{\nu},~~~~\x\equiv |\vc{\x}|.
 \end{equation}

 An important property of the ADHM construction is that operator
 $\L^{\dagger}(x)\L(x)$ is scalar, i.e. proportional to the unit matrix,
 and positive definite. We shall need an explicit expression for it's
 Green function \cite{Nahm80}:
 \begin{eqnarray}\label{f}
 f(x;z,z')&\equiv&
 \langle z|(\L^{\dagger}(x)\L(x))^{-1}|z'\rangle \nonumber\\
 &=&-\frac{1}{2\x}\E^{i\x_0(z-z')}
 \left(\sinh\x |z-z'|+\coth\frac{\x}{2}\,\sinh\x z\,\sinh\x z'
 \right.\nonumber\\&&\left.
 -\tanh\frac{\x}{2}\,\cosh\x z\,\cosh\x z'\right).
 \end{eqnarray}

 The isospin--$1$ Green function of $\left(-D^2\right)$ is expressed
  through
 $|v\rangle$ in a relatively simple way \cite{CWS}:
 \begin{eqnarray}\label{green1}
 G^{ab}(x,y)&=&\frac
 {\frac12\tr\s^a\langle v(x)|v(y)\rangle
 \s^b\langle v(y)|v(x)\rangle}
 {4\pi^2(x-y)^2} \nonumber\\
 &&+\frac{1}{4\pi^2}\int_{-1/2}^{1/2}dz_1\,dz_2\,dz_3\,dz_4\,
 M(z_1,z_2,z_3,z_4)     \nonumber\\
 &&\times\frac12\tr\left(v^{\dagger}(x,z_1)
 v(x,z_2)\s^a\right)
 \tr\left(v^{\dagger}(y,z_4)v(y,z_3)\s^b\right),
 \end{eqnarray}
 where \cite{Nahm80}
 \begin{eqnarray}\label{kern1}
 M(z_1,z_2,z_3,z_4)&=&-\,\frac{\m^2}{4}\,\D(z_1-z_2-z_3+z_4)
 \left[|z_1+z_2-z_3-z_4|-1\right.
 \nonumber\\
 &&\left.+|z_1-z_2|+\frac{(z_1+z_2)(z_3+z_4)}
 {1-|z_1-z_2|}\right].
 \end{eqnarray}
 This Green function does not obey a periodicity condition. The
 periodic one is obtained from it by standard procedure:
 \begin{equation}\label{green}
 \G^{ab}(x,y)=\sum_{n=-\infty}^{+\infty}G^{ab}(x_0,{\bf x};
 y_0+n\b,{\bf y}).
 \end{equation}

 It is
 convenient to represent regularized Green function \rf{grdef} as a sum
 of three terms:  \begin{equation}\label{greenr}
 \G_{R}^{ab}(x,y)=\G^{(1)ab}(x,y)+\G^{(2)ab}(x,y)+\G^{(3)ab}(x,y),
 \end{equation}
 where
 \begin{eqnarray}\label{g1}
 \G^{(1)ab}(x,y)&=&\frac{\frac12
 \tr\left(\s^a \langle v(x)|v(y)\rangle\s^b\langle v(y)|v(x)\rangle
 -\s^a\Phi(x,y)\s^b\Phi(y,x)\right)}{4\pi^2(x-y)^2} \nonumber\\
 &=&\frac{1}{4\pi^2}\frac12
 \tr\left(\s^aR(x,y)\s^b\Phi(y,x)+\s^a\Phi(x,y)\s^bR(y,x)\right)
 \nonumber\\&&+O\left((x-y)^2\right),
 \end{eqnarray}
 \begin{equation}\label{fr}
 R(x,y)=\frac{\langle v(x)|v(y)\rangle-\Phi(x,y)}{(x-y)^2},
 \end{equation}
 \begin{equation}\label{g2}
 \G^{(2)ab}(x,y)=\sum_{n\neq 0}\frac
 {\frac12\tr\s^a\left\langle v(x)\left|\E^{2\pi iknz}
 \right|v(y)\right\rangle\s^b
 \left\langle v(y)\left|\E^{-2\pi iknz}
 \right|v(x)\right\rangle}
 {4\pi^2\left[(x_0-y_0-n\b)^2+({\bf x}-{\bf y})^2\right]}.
 \end{equation}
 In the last expression it was taken into account that
 \begin{equation}\label{vn}
 \left|v(y_0+n\b,{\bf y})\right\rangle
 =\E^{2\pi iknz}\left|v(y_0,{\bf y})\right\rangle,
 \end{equation}
 where $k$ is defined by \rf{mk}. Here and in what follows we consider
 $k$ as a continuous variable, and put it to be integer only at the end
 of the calculation. Using eq. \rf{vn} one can transform
 $\G^{(3)}$, which
 comes from the second term in \rf{green1}, into a more simple form.
 Really, summation over $n$ in \rf{green} gives rise to a factor
 \begin{equation}\label{fourier} \sum_n \E^{2\pi
 ikn(z_3-z_4)}=\frac{1}{k}\sum_n\D\left(z_3-z_4- \frac{n}{k} \right),
 \end{equation} which together with the $\D$--function in the kernel
 \rf{kern1} allows to eliminate an integration over $z_2$ and $z_3$ in eq.
 \rf{green1}.  Changing variables in the remaining integrals to
 $z=2z_1-n/k$ and $w=2z_4+n/k$ one finally gets:
 \begin{eqnarray}\label{g3} \G^{(3)ab}(x,y)&=&-\frac{\m^2}{16 \pi^2 k}
 \sum_{n=-N}^{N}\frac14
 \int_{-1+|n|/k}^{1-|n|/k}dz\,dw\,\left(|z-w|-1+\frac{|n|}{k}
 +\frac{zw}{1-\frac{|n|}{k}}\right) \nonumber\\
 &&\times\frac12\tr(u_{-n}(x,z)\s^a)
 \tr(u_{n}(y,w)\s^b),
 \end{eqnarray}
 where
 \begin{equation}\label{un}
 u_n(x,z)=\frac{\x}{\sinh\x}\exp\left(i\x_0\frac{n}{k}-\x_i\s^iz\right)
 \end{equation}
 and $N$ is an integral part of $k$. In \rf{g3} $n$ varies from $-N$ to
 $N$, because in \rf{fourier} $|z_3-z_4|$ is by definition smaller than
 unity.

 The next step is evaluation of the current $\J_{\m}$, but first we
 rescale integration variables in eq. \rf{vardet} and express the trace in
 the adjoint representation in it through the usual matrix trace:
 \begin{equation}\label{var} k\frac{\d\g_I}{\d
 k}=\int_0^{2\pi k}d\x_0\, \int d^3\x\,\tr
 \left(\m\frac{\d A_{\nu}}{\d\m}J_{\nu}\right)
 =8\pi^2k\int_0^{\l k}d\x\,\x^2
 \tr
 \left[\left(\x_jF_{j\nu}+\D_{\nu 0}A_0\right)J_{\nu}\right],
 \end{equation}
 where $\l=2\pi R/\b$, $R$ is an infrared cutoff.   The last equality
 holds, because an integrand depends only on $|\vc{\x}|$ in virtue of the
 central symmetry and time independence of the monopole solution.

 Consider first the contribution of $\G^{(1)}$. It can be rewritten
 in the
 form \rf{var} with the help of the identities
 \begin{eqnarray}\label{id}
 \dl^{ad}\frac12\tr
 \left(\s^dA\s^bB\right)=\frac12\tr\left[\s^a(\dl A)\s^bB-
 \s^aA\s^b(B\dr)\right], \nonumber\\
 \frac12\tr
 \left(\s^aA\s^dB\right)\dr^{db}=\frac12\tr\left[\s^a(A\dr)\s^bB-
 \s^aA\s^b(\dl B)\right],
 \end{eqnarray}
 where the covariant derivative on the right hand side acts in the
 fundamental representation. Applying them to \rf{g1} and taking into
 account that
 \[
 \left.\dl \Phi(x,y)\right|_{y=x}=0=\left.\Phi(y,x)\dr\right|_{y=x},
 \]
 one finds that $\J^{(1)ba}_{\m}=\frac14i\e^{abd}
 \tr\left(\s^dJ^{(1)}_{\m}\right)$ with
 \begin{equation}\label{j1}
 J^{(1)}_{\m}(x)=\left.\frac{1}{\pi^2}\left(\dl R(x,y)
 +R(y,x)\dr\right)\right|_{y=x}.
 \end{equation}

 Now one  may use the ADHM construction to write \cite{CGOT}:
 \begin{equation}\label{mainj1}
 J_{\m}^{(1)}=\frac{1}{3\pi^2}
 \left\langle v\left|f
 \left(e_{\m}\L^{\dagger}-\L e_{\m}^{\dagger}\right)
 f\right|v\right\rangle\,,~~~~e_{\m}=(1,-i\vc{\s}).
 \end{equation}
 From \rf{v} and \rf{f} we find
 \begin{equation}\label{fv}
 f|v\rangle=-\frac12(\x\sinh\x)^{-1/2}\E^{i\x_0z}(a-n_i\s^ib),
 \end{equation}
 \begin{equation}\label{defa}
 a=z\sinh\x z-\frac12\tanh\frac{\x}{2}\,\cosh\x z,
 \end{equation}
 \begin{equation}\label{defb}
 b=z\cosh\x z-\frac12\coth\frac{\x}{2}\,\sinh\x z.
 \end{equation}
 Substituting \rf{fv} in eq. \rf{mainj1}
  one obtains:
  \begin{eqnarray}\label{j1t}
 J_{0}^{(1)}&=&\frac{1}{6\pi^2\sinh\x}\int^{1/2}_{-1/2}dz\,
 [in_i\s^i(a^2+b^2)-2iab]
 \nonumber\\
 &=&in_i\s^i\frac{1}{12\pi^2}
 \left(\frac{1}{\x^3}-\frac{\cosh\x}{\sinh^3\x}\right),
 \\
 \label{j1s}
 J_{i}^{(1)}&=&\frac{1}{6\pi^2\x\sinh\x}\int_{-1/2}^{1/2}dz\,
 \left\{i\e_{ijk}n_j\s^k\left[\x(a^2-b^2)
 +b\frac{da}{dz}
 -a\frac{db}{dz}\right]
 \right.\nonumber\\
 &&\left.+\s^ia\frac{da}{dz}+n_jn_k\s^j\s^i\s^k
 b\frac{db}{dz}-n_i\left(a\frac{db}{dz}+b\frac{da}{dz}\right)\right\}
 \nonumber\\
 &=&i\e_{ijk}n_j\s^k\frac{1}{12\pi^2\sinh\x}\left(\frac{1}{\sinh^2\x}
 +\frac{\cosh\x}{\x\sinh\x}-\frac{2}{\x^2}\right).
 \end{eqnarray}
 For the first factor in the integrand in eq. \rf{var} we have:
 \begin{equation}\label{varft}
 \x_jF_{j0}+A_0=\frac{\s^an_a}{2i}\left(\coth\x-\frac{\x}{\sinh^2\x}
 \right)
 \end{equation}
 \begin{equation}\label{varfs}
 \x_jF_{ji}=\e_{ija}n_j \frac{\s^a}{2i}\frac{\x}{\sinh\x}
 \left(\coth\x-\frac{1}{\x}\right).
 \end{equation}
 Calculating the trace and doing the integral over $\x$ one finds:
 \begin{equation}\label{gm1}
 k\frac{\g^{(1)}_I}{\d k}=\left(\frac23\ln\frac{\l
 k}{\pi}-\frac{5}{18}+\frac{\pi^2}{54}+\frac{2\gamma}{3}\right)\,k,
 \end{equation}
 where $\gamma$ is the Euler constant. Note that the term containing
 $J_{0}^{(1)}$ is logarithmically divergent -- this is the origin of
 $\ln\l$ in \rf{gm1}.

 Let us turn to the contribution of $\G^{(2)}$. To rewrite it in the form
 \rf{var} we use \rf{id} and an obvious equality
 $D_{\m}=\langle v(x)|\d_{\m}|v(x)\rangle$, valid for covariant derivative
 in the fundamental representation. A simple calculation gives:
 \begin{eqnarray}\label{j2ab}
 \J_{\m}^{(2)ab}&=&\frac{1}{4\pi^2}\sum_{n\neq 0}\frac12\tr
 \left(\D_{\m 0}\frac{4}{\b^3 n^3}
 \s^a\left\langle v\left|\E^{2\pi iknz}
 \right|v\right\rangle\s^b
 \left\langle v\left|\E^{-2\pi iknz}
 \right|v\right\rangle
 \right.\nonumber\\
 &&+\frac{1}{\b^2 n^2}
 \s^a\left\langle v\left|\left[\d_{\m}\hat{P},\E^{2\pi iknz}\right]
 \right|v\right\rangle\s^b
 \left\langle v\left|\E^{-2\pi iknz}
 \right|v\right\rangle
 \nonumber\\
 &&\left.+\frac{1}{\b^2 n^2}
 \s^a\left\langle v\left|\E^{2\pi iknz}
 \right|v\right\rangle
 \s^b\left\langle v\left|\left[\E^{-2\pi iknz},\d_{\m}\hat{P}\right]
 \right|v\right\rangle
 \right).
 \end{eqnarray}
 with $\hat{P}=|v\rangle\langle v|$. Substituting this expression in
 \rf{vardet} we rewrite it in the form \rf{var} with
 \begin{eqnarray}\label{mainj2}
 J_{\m}^{(2)}&=&\frac{1}{4\pi^2}\sum_{n\neq 0}
 \left[
 \frac{1}{(2\pi k)^2 n^2}
 \left(
 \left\langle v\left|\left[\d_{\m}\hat{P},\E^{2\pi iknz}\right]
 \right|v\right\rangle
 \tr\left\langle v\left|\E^{-2\pi iknz}
 \right|v\right\rangle \right.\right.\nonumber\\
 &&\left.-\left\langle v\left|\E^{-2\pi iknz}
 \right|v\right\rangle
 \tr\left\langle v\left|\left[\d_{\m}\hat{P},\E^{2\pi iknz}\right]
 \right|v\right\rangle+(n\leftrightarrow -n)
 \right)   \nonumber\\
  &&+\left.
  4\D_{\m 0}\frac{1}{(2\pi k)^3 n^3}
 \left(
 \left\langle v\left|\E^{2\pi iknz}
 \right|v\right\rangle
 \tr\left\langle v\left|\E^{-2\pi iknz}
 \right|v\right\rangle
 -(n\leftrightarrow -n)
 \right)
 \right].
 \end{eqnarray}
 At first sight $J^{(2)}$, being of order $1/k^2$,  gives $O(1/k)$
 contribution   to $\g$. However,
 the situation is more involved due to infrared divergences. The
 calculations are lengthy and are carried out in Appendix. Here we only
 quote the results. The leading cubic divergence has the form
 \begin{equation}\label{irdiv} \ln\det\left(-D^2\right)\sim
 \frac{16\pi^3R^3}{9\b^3}s^2(1-s)^2=\frac{4\pi^2}{3} VT^3s^2(1-s)^2,
 \end{equation}
 where $s=k\,\mod\,1$. Hence we obtain the expected result, that the
 contribution of the monopoles with nonintegral topological charge is
 suppressed by a factor $\exp(-cVT^3)$ with coefficient $c$ precisely the
 same as for the constant background field \cite{GPY}. For integral  $k$
 the cubic divergence vanishes, all other ones also cancel, and
 $\g_{I}^{(2)}$ is infrared finite and is
 indeed of order $1/k$.

 Now let us proceed with the contribution of $\G^{(3)}$. Direct
 calculation of $\J^{(3)}$ according to \rf{defj} and it's substitution
 into \rf{vardet} leads, after a simple algebra, to the result, which
 again has a form of \rf{var} with
 \begin{eqnarray}\label{j3}
 J^{(3)}_{\m}&=&\frac{1}{32\pi^2 k}\sum_{n=-N}^{N}
 \int_{-1+|n|/k}^{1-|n|/k}dz\,dw\,\left(|z-w|-1+\frac{|n|}{k}
 +\frac{zw}{1-\frac{|n|}{k}}\right)
 \nonumber\\
 &&
 \left\{
 \frac12\left[u_{-n}(z),\d_{\m}u_n(w)\right]
 -\frac12\left[\d_{\m}u_{-n}(z),u_{n}(w)\right]
 \right.\nonumber\\
 &&+u_{-n}(z)\tr\left(u_n(w)A_{\m}\right)
 +u_{n}(w)\tr\left(u_{-n}(z)A_{\m}\right)
 \nonumber\\
 &&\left.-2A_{\m}\tr\left(u_{-n}(z)u_n(w)\right)
 +A_{\m}\tr u_{-n}(z)\tr u_n(w)
 \right\}.
 \end{eqnarray}
 Note that the term under consideration do not contain infrared
 divergences, because $J_0^{(3)}=0$. It can be shown by an explicit
 calculation according to eq.  \rf{j3},  but more simple way to see it
 is to use directly a definition \rf{defj} and eq. \rf{g3}. Really, since
 $A_0^c\propto n^c$ and $\tr\left(u_n(x,z)\s^d\right)\propto n^d$,
 $\e^{acb}A_0^c\tr\left(u_n(x,z)\s^d\right)=0$ and $D_0^{ad}$ acts on
 $\G^{(3)db}$ as an ordinary time derivative. Hence $\J_0^{ab}\propto
 \tr(u_{-n}\s^a)\tr(u_{n}\s^b)\propto n^an^b$, which gives zero after the
  substitution in eq. \rf{vardet}.

 The straightforward calculation of $J_i^{(3)}$ gives:
 \begin{eqnarray}\label{mainj3}
 J_i^{(3)}&=&-i\e_{ijk}n_j\s^k\frac{1}{16\pi^2 k}\frac{\x^2}{\sinh^3\x}
 \sum_{n=-N}^{N}
 \int_{-1+|n|/k}^{1-|n|/k}dz\,dw\,\left(|z-w|-1+\frac{|n|}{k}
 \right.\nonumber\\
 &&\left.+\frac{zw}{1-\frac{|n|}{k}}\right)
 \sinh\x z\sinh\x w \nonumber\\
 &=&i\e_{ijk}n_j\s^k\frac{1}{16\pi^2 k}\frac{\x^2}{\sinh^3\x}
 \sum_{n=-N}^{N}
 \left[
 \frac{2}{\x^2}\left(1-\frac{|n|}{k}\right)+\frac{1}{\x^3}\sinh
 2\left(1-\frac{|n|}{k}\right)\x
 \right.\nonumber\\
 &&\left.-\frac{2}{\x^4}
 \frac{\cosh 2\left(1-\frac{|n|}{k}\right)\x-1}{1-\frac{|n|}{k}}
 \right]
 \nonumber\\
 &=&i\e_{ijk}n_j\s^k\frac{1}{16\pi^2 k}\frac{\x^2}{\sinh^3\x}
 \left[
 \frac{2}{\x^2}+\frac{\sinh 2\x}{\x^3}-\frac{2}{\x^4}(\cosh 2\x-1)
 \right.\nonumber\\
 &&+\frac{4}{\x^2}N\left(1-\frac{N+1}{2k}\right)
 +\frac{1}{\x^3}
 \frac{\cosh\left(2-\frac1k\right)\x-\cosh(2s-1)\frac{\x}{k}}
 {\sinh\frac{\x}{k}}
 \nonumber\\
 &&\left.-\frac{4}{\x^4}\int_{0}^{\x}d\eta\,
 \frac{\cosh\left(2-\frac1k\right)\eta -\cosh(2s-1)\frac{\eta}{k}}
 {\sinh\frac{\eta}{k}}
 \right].
 \end{eqnarray}
 After the substitution of this expression in eq. \rf{var} the resulting
 integral over $\x$  cannot be expressed in elementary functions.
 But the $k\rightarrow\infty$ asymptotics can be found explicitly.
 Since the integral is infrared convergent,
 we may simply expand two last
 terms in \rf{mainj3} in powers of $1/k$, which
 gives, after integration:  \begin{equation}\label{gm3}
 k\frac{\g^{(3)}_I}{\d k}=\left(\frac13+\frac{\pi^2}{9}\right)
 N\left(1-\frac{N+1}{2k}\right)
 +(3-2\ln
 2\pi)k+\left(\frac16+\frac{\pi^2}{18}\right)+O\left(\frac1k\right).
 \end{equation}

 This completes the calculation of $\d\g_I/\d k$. The result is given by
 \rf{gm1} and \rf{gm3}; $\g^{(2)}_I$  gives $O(1/k)$ contribution, as it
 is shown in Appendix. To obtain $\g_I$
 one should remove the variation with respect to $k$. An integration of
 $\d\g_I^{(1)}/\d k$ and of the last two terms in $\d\g_I^{(3)}/\d k$ is
 trivial.  As for the first term in $\d\g_I^{(3)}/\d k$,
   it is convenient
 to divide the integration region in the integral intervals, i.e. to
 integrate first over $s=k\,\mod\,1$ from $0$ to $1$ at fixed $N\equiv
 k-s$ and then to sum over $N$:  \begin{eqnarray}\label{}
 \sum_{N<k}\int_0^1ds\,\frac{N}{N+s}\left(1-\frac{N+1}{2}\frac{1}{N+s}
 \right)&=&\sum_{N<k}\left(\frac12-\frac{1}{2N}+O\left(\frac
 {1}{N^2}\right)\right) \nonumber
 \\&=&\frac12k-\frac12\ln k+O\left(k^0\right).\nonumber
 \end{eqnarray}
 Note that the term, proportional to $\ln k$, cancels with the result of
 the integration of the last one in \rf{gm3}. Adding up all
 contributions we obtain:
 \begin{equation}\label{lndet} \g_I=\frac23k\ln
 k+\left(\frac23\ln\frac{2\pi R}{\b}
 +\frac{20}{9} +\frac{2\pi^2}{27}-2\ln
 2-\frac{8}{3}\ln\pi+\frac{2\gamma}{3}\right)k+O\left(k^0\right).
 \end{equation}

 \newsection{Ultraviolet divergent part of determinant}

 In this section the surface term \rf{ga2} is considered. It contains,
 since $\langle x|(-D^2)^{-2}|y\rangle\sim\frac{1}{16\pi^2}\ln(x-y)^2$,
 the logarithmic ultraviolet divergence of the form $-\frac23k\ln\r$. This
 is the conventional contribution of a scalar loop to the coupling
 constant renormalization. To calculate $\g_{II}$ completely we define
 \begin{equation}\label{defs}
 S^{ab}(\x,\x')=\int d^4\a\,G^{ac}(\x,\a)G^{cb}(\a,\x').
 \end{equation}
 where the integration ranges over the whole Euclidean space and the
 rescaled variables \rf{x} are used. The squared propagator,
 $(-D^2)^{-2}$, which enters eq. \rf{ga2}, can be obtained from $S$ by the
 same procedure as the periodic Green function was obtained from the
 nonperiodic one in \rf{green}. But, since $S(\x,\x')$ depends only on the
 difference $\x'_0-\x_0$, it is more convenient to Fourier transform in
 $\x_0$. In such momentum--coordinate representation the asymptotics of
 the heat kernel for large $\r$ has the form
 \begin{equation}\label{ahk}
 \E^{D^2/\r^2}\sim\D({\bf x}-{\bf y})\E^{-\o^2\m^2/\r^2}
 \end{equation}
 and, as
 \begin{equation}\label{aaf}
 \left(\m\frac{\d A_{0}}{\d\m}F_{i0}\right)^{ab}=\frac{\m}{R^2}n_i
 (n_an_b-\D_{ab})+O\left(\E^{-\m R}\right),
 \end{equation}
 we rewrite eq. \rf{ga2} as follows:
 \begin{equation}\label{mainga2}
 k\frac{\d\g_{II}}{\d k}=\frac13\oint_{S_R}d^2\Sigma\,\frac{1}{R^2}
 \sum_{n=-\infty}^{+\infty}\phi(\o_n,\vc{\x})\E^{-\o_n^2\m^2/\r^2},
 ~~~~\o_n=\frac{n}{k},
 \end{equation}
 where
 \begin{equation}\label{defphi}
 \phi(\o,\vc{\x})=\int_{-\infty}^{+\infty}dt\,\E^{-i\o t}(n_an_b-\D_{ab})
 S^{ab}(0,\vc{\x};t,\vc{\x}).
 \end{equation}
 This function can be evaluated using the explicit form of the propagator
 \rf{green1}. Note that the second term in eq. \rf{green1} is proportional
 to $n_an_b$ and thus does not contribute to $\phi(\o,\vc{\x})$. The
 integration over $\a_0$ in \rf{defs} and over $t$ in \rf{defphi} can be
 easily done, which yields:
 \begin{eqnarray}\label{ga20}
 \phi(\o,\vc{\x})&=&\int d^3\a\,\frac{1}
 {64\pi^2\left(\vc{\x}-\vc{\a}\right)^2}
 (n_an_b-\D_{ab})
 \left(2\tr
 \left\langle v\left|\hat{Q}^{a}(\o)\right|v\right\rangle
 \left\langle v\left|\hat{Q}^{b}(-\o)\right|v\right\rangle
 \right.
 \nonumber\\&&\left.
 -\tr\left\langle v\left|\hat{Q}^{a}(\o)\right|v\right\rangle\,
 \tr\left\langle v\left|\hat{Q}^{b}(-\o)\right|v\right\rangle
 \right),
 \end{eqnarray}
 where $|v\rangle\equiv|v(0,\vc{\a})\rangle$ and
 \begin{equation}\label{defqa}
 \left\langle z\left|\hat{Q}^{a}(\o)\right|w\right\rangle=
 v(0,\vc{\x};z)\s^av^{\dagger}(0,\vc{\x};w)\E^{-|z-w+\o|\,|\vc{\x}
 -\vc{\a}|}.
 \end{equation}
 The integral \rf{ga20} must be evaluated for $\x\rightarrow\infty$. In
 this limit only the domain of integration with $\a\sim\x$ gives essential
 contribution. In this region $v(0,\vc{\a};z)$ and $v(0,\vc{\x};z)$ are
 exponentially small, unless $z$ is closed to $\pm 1/2$, when
 $v(0,\vc{\a};z)\sim\frac12\sqrt{\a}\E^{(1/2\mp z)\a}(1\mp\s^i\nu_i)$ and
 $v(0,\vc{\x};z)\sim\frac12\sqrt{\x}\E^{(1/2\mp z)\x}(1\mp\s^in_i)$, where
 $\nu_i=\a_i/\a$ and $n_i=\x_i/\x$. Then, with exponential accuracy,
 \begin{eqnarray}\label{qa}
 \left\langle v\left|\hat{Q}^{a}(\o)\right|v\right\rangle&=&
 \frac14\x\a\sum_{p,q=\pm 1}\int_{-1/2}dz\,dw\,\E^{(1-z-w)(\x+\a)-
 |pz+qw+\o|\,|\vc{\x}-\vc{\a}|}\nonumber\\
 &&\times(1+p\s^i\nu_i)(1+p\s^jn_j)\s^a(1-q\s^kn_k)(1-q\s^l\nu_l).
 \end{eqnarray}
 The substitution of this expression for $\left\langle
 v\left|\hat{Q}^{a}(\o)\right|v\right\rangle$ in eq. \rf{ga20}
 considerably simplifies the matrix algebra, which can be performed with
 the use of the identities
 \[
 (1+p\s^in_i)\s^jn_j=\s^jn_j(1+p\s^in_i)=p(1+p\s^in_i),\]\[
 (1+p\s^in_i)(1+q\s^jn_j)=2\D_{pq}(1+p\s^in_i).
 \]
 After some calculations the integral \rf{ga20} can be rewritten
  in the
 following compact form:
 \begin{equation}\label{ga21}
 \phi(\o,\vc{\x})=-\frac{1}{4\pi^2}\int d^3\a\,\x^2\a^2(1+n_i\nu_i)^2
 \left(M^2(\o)+M^2(-\o)\right),
 \end{equation}
 where
 \begin{equation}\label{defio}
 M(\o)=\int_{-1/2}dz\,dw\,\E^{(1-z-w)(\x+\a)-
 |z+w-\o|\,|\vc{\x}-\vc{\a}|}.
 \end{equation}

 It is convenient to change the integration variables in \rf{ga21} to
 \begin{equation}\label{newv}
 x=\x+\a,~~~~y=|\vc{\x}-\vc{\a}|,
 \end{equation}
 then
 \begin{equation}\label{ga22}
 \phi(\o,\vc{\x})=-\frac{1}{8\pi\x}\int_{0}^{\infty}\frac{dy}{y}\,
 \int_{\x+|\x-y|}^{2\x+y}dx\,(x-\x)(x^2-y^2)^2
 \left(M^2(\o)+M^2(-\o)\right).
 \end{equation}
 Assuming that an upper bound of integration in \rf{defio} is equal to
 infinity, one finds:
 \begin{equation}\label{io}
 M(\o)=\left\{\begin{array}{ll}{\displaystyle
 \frac{\E^{(\o-1)y}}{(x+y)^2},} & {\displaystyle\o\leq 1}\\
 {\displaystyle\frac{\E^{-(\o-1)y}}{(x-y)^2}-\E^{-(\o-1)x}
 \left[\frac{2(\o-1)y}{x^2-y^2}+\frac{4xy}{(x^2-y^2)^2}\right],}
 &{\displaystyle \o\geq 1.}\end{array}\right.
 \end{equation}
 When $|\o|\neq 1$, the integrand in \rf{ga22} is exponentially small for
 $y\sim\x$, so one can restrict the integration region to $y<<\x$. Then
 the integral can be easily calculated and one obtains, up to $O(1/\x)$
 terms,
 \begin{equation}\label{ga23}
 \phi(\o,\vc{\x})=\left\{\begin{array}{ll}{\displaystyle
 -\frac{1}{4\pi(1-\o^2)},} & {\displaystyle|\o|< 1}\\
 {\displaystyle -\frac{|\o|}{4\pi(\o^2-1)},}
 &{\displaystyle |\o|> 1.}\end{array}\right.
 \end{equation}
 This expression is singular at $\o=1$. Remind that in the dimensionless
 units we use the size of the monopole is equal to unity. In fact, eq.
 \rf{ga23} is not valid in the resonant case. The point is  that, when
 $|\o^2-1|\<< 1/\x$, one can not restrict the integration region to
 $y<<\x$. In particular, $\phi(\pm 1, \vc{\x})$ is finite of order $\x$.
 The general expression for $\phi$, valid also for resonant frequencies, is
 too cumbersome, but in what follows we shall need only the following
 integrals:
 \begin{equation}\label{def+-}
 I_-(\dq,\vc{\x})=\int_{1-\dq}^{1}d\o\,\phi(\o,\vc{\x}),~~~~
 I_+(\dq,\vc{\x})=\int_{1}^{1+\dq}d\o\,\phi(\o,\vc{\x}).
 \end{equation}
 For $1>>\dq>>1/\x$ one gets:
 \begin{equation}\label{i-}
 I_-(\dq,\vc{\x})=-\frac{1}{8\pi}\left(\ln\x\dq-\frac52+
 \frac{\pi^2}{3}-7\ln 2+\gamma\right),
 \end{equation}
 \begin{equation}\label{i+}
 I_+(\dq,\vc{\x})=-\frac{1}{8\pi}\left(\ln\x\dq+\frac32
 +\ln 2+\gamma\right).
 \end{equation}

 Returning to the eq. \rf{mainga2} we find that, when the fractional part
 of $k$ is not too close to $0$ or to $1$, say $N\dq<s<1-(N+1)\dq$, eq.
 \rf{ga23} can be used for $\phi(\o_n,\vc{\x})$. On the other hand, when
 $s<N\dq$ or $s>1-(N+1)\dq$, the sum entering \rf{mainga2} can be
 approximated by $2\phi(\o_N,\vc{\x})$ or by $2\phi(\o_{N+1},\vc{\x})$,
 respectively. By noting that
 \[
 \int_{0}^{N\dq}\frac{ds}{k}\,\phi(\o_N,\vc{\x})=I_-(\dq,2\pi NT{\bf x}),
 \]\[
 \int_{1-(N+1)\dq}^{1}\frac{ds}{k}\,\phi(\o_{N+1},\vc{\x})=I_+(\dq,2\pi
 (N+1)T{\bf x}),
 \]
 we find:
 \begin{eqnarray}\label{ga2f}
 \g_{II}&=&\frac13\int^k dk'\,\frac{4\pi}{k'}\sum_{n=-\infty}^{+\infty}
 \phi(\o_n,\vc{\x})\E^{-\o_n^2\m^2/\r^2}
 \nonumber\\&=&\frac13\sum_{N<k}\left\{
 8\pi I_-(\dq,2\pi NT{\bf x})+
 8\pi I_+(\dq,2\pi (N+1)T{\bf x})\right.
 \nonumber\\
 &&-\int_{N\dq}^{1-(N+1)\dq}ds\,\left[
 2\ln\frac{\r}{2\pi T}+2\psi(2N+s+1)-2\psi(s)
 \right.\nonumber\\&&\left.\left.-2\psi(N+1)-\pi\cot\pi
 (N+s)-\gamma
 \right]\right\}
 \nonumber\\
 &=&-\frac13k\left(2\ln R\r-1+\frac{\pi^2}{3}-4\ln 2+\gamma\right)
 +O\left(k^0\right).
 \end{eqnarray}

 Collecting together all contributions, given by eqs. \rf{ga3main},
 \rf{lndet}  and \rf{ga2f}, we obtain the final result, quoted in sec.
 \ref{mnpl}:
 \begin{equation}\label{gaf}
 \g=\frac23k\ln\frac{T}{\r}+\frac{2}{3}k \ln k+
 \left(\frac{20}{9}-\frac{\pi^2}{27}-2\ln \pi+\frac{\gamma}{3}\right)k
 +O\left(k^0\right).
 \end{equation}

 \newsection{Conclusions}

 Our results show that, despite the Coulomb nature of the monopole field,
 the one--loop corrections to the action are free from the infrared
 divergences. Thus the contribution of the BPS monopoles can, in
 principle, be calculated by semiclassical methods. To perform this
 calculation completely it is important to consider zero modes in the
 monopole background and related nonstatic deformations of the classical
 solution. The number of the zero modes grows with the decrease of the
 monopole size, which makes the problem of the calculation of the monopole
 density more complicated, than that for the instantons.

 It is worth mentioning that generalization to $SU(N_c)$ with $N_c>2$ and
 introduction of massless fermions also should not be
 literally analogous to
 that in the instanton calculations, because of a difference in the
 structure of gauge zero modes and the fact that the topological charge of
 the BPS monopole, and hence the number of fermion zero modes, is related
 to the monopole size. In particular, the form of the 't~Hooft interaction
 \cite{tHooft}, induced by the BPS monopoles, may be not the same as for
 the instantons.

 \newsection*{Acknowledgments}

The author is grateful to A.~Abrikosov~(Jr.), E.~Akhmedov, E.~Gubankova
and K.~Selivanov for useful discussions. The work was supported in part
 by ISF grant MET~000, by INTAS grant 94-0840 and by RFFR grant
 94-01-00285.

 \setcounter{section}{0}
 \setcounter{subsection}{0}

 \appendix{Calculation of $\g_{I}^{(2)}$}

 In this appendix we calculate $\g_{I}^{(2)}$, the variation of which
 is given by \rf{var} with $J^{(2)}$ defined in \rf{mainj2}. First, we
 note that \begin{eqnarray}\label{com} \left\langle
 v\left|\left[\d_{\m}\hat{P},\E^{2\pi iknz}\right] \right|v\right\rangle
 &=&\left\langle \d_{\m}v\left|\E^{2\pi iknz}
 \right|v\right\rangle
 -\left\langle v\left|\E^{2\pi iknz}
 \right|\d_{\m}v\right\rangle
 \nonumber\\
 &&+\left\{A_{\m},\left\langle v\left|\E^{2\pi iknz}
 \right|v\right\rangle\right\}.
 \end{eqnarray}

 We begin with the term,  containing $J_0^{(2)}$.
 Define \begin{equation}\label{en} E_n\equiv\left\langle v\left|\E^{2\pi
 iknz} \right|v\right\rangle=C_n-in_i\s^iD_n,
 \end{equation} \begin{equation}\label{cn}
 C_n=\frac{\x^2\cos\pi k n+\x\coth\x \,\pi kn\sin\pi kn}
 {\x^2+\pi^2k^2n^2},
 \end{equation}
 \begin{equation}\label{dn}
 D_n=\frac{\x^2\coth\x\,\sin\pi k n-\x\pi kn\cos\pi kn}
 {\x^2+\pi^2k^2n^2},
 \end{equation}
 then, since $\left|\d_0v\right\rangle=iz\left|v\right\rangle$,
 \begin{equation}\label{com0}
 \left\langle v\left|\left[\d_{0}\hat{P},\E^{2\pi iknz}\right]
 \right|v\right\rangle
 =-\frac{1}{\pi n}\frac{\d E_n}{\d k}+\left\{A_0,E_n\right\}.
 \end{equation}
 Substituting this expression into \rf{mainj2} one gets:
 \begin{eqnarray}
 \label{j02}
 J_{0}^{(2)}&=&\frac{1}{8\pi^2}\sum_{n\neq 0}
 \left[
 \frac{1}{\pi^3k^3n^3}\left(1-\frac12k\frac{\d}{\d k}\right)
 \left(E_n\tr E_{-n}-E_{-n}\tr E_n\right)
 \right.\nonumber\\
 &&+\frac{1}{2\pi^2k^2n^2}
 \left\{A_0,E_n\tr E_{-n}+E_{-n}\tr E_n\right\}
 \nonumber\\
 &&\left.+\frac{1}{\pi^2k^2n^2}E_{-n}\tr\left(A_0E_{n}\right)
 -\frac{1}{\pi^2k^2n^2}E_{n}\tr\left(A_0E_{-n}\right)
 \right].
 \end{eqnarray}
 Using \rf{en} and taking into account that
 \[
 \frac{1}{k^3}\left(1-\frac12k\frac{\d}{\d k}\right)=-\frac12
 \frac{\d}{\d k}\frac{1}{k^2},
 \]
 we find:
 \begin{eqnarray}\label{mainj02}
 J_{0}^{(2)}&=&\frac{1}{2\pi^2}
 \left[
 in_i\s^i\frac12\frac{\d}{\d k}\left(\frac{1}{\pi^3k^2}
 \sum_{n\neq 0}\frac{C_nD_n}{n^3}\right)
 +A_0\frac{1}{\pi^2k^2}\sum_{n\neq 0}\frac{C_n^2-D_n^2}{n^2}
 \right]
 \nonumber\\
 &=&in_i\s^i\frac{1}{4\pi^2}\left[\frac{\d P}{\d k}
 -\left(\coth\x-\frac{1}{\x}\right)Q\right],
 \end{eqnarray}
 where
 \begin{eqnarray}\label{defp}
 P=\frac{1}{\pi^3k^2}\sum_{n\neq 0}\frac{C_nD_n}{n^3}
 &=&\frac{1}{2\pi^3k^2}
 \left[
 \x^4\coth\x\sum_{n\neq 0}\frac{\sin2\pi kn}
 {n^3\left(\x^2+\pi^2k^2n^2\right)^2}
 \right.\nonumber\\ &&
 -\pi^2k^2\x^2\coth\x\sum_{n\neq 0}\frac{\sin2\pi kn}
 {n\left(\x^2+\pi^2k^2n^2\right)^2}
 \nonumber\\ &&
 -\pi k\frac{\x^3\cosh 2 \x}{\sinh^2\x}
 \sum_{n\neq 0}\frac{\cos2\pi kn}
 {n^2\left(\x^2+\pi^2k^2n^2\right)^2}
 \nonumber\\&&\left.
 +\pi k\frac{\x^3}{\sinh^2\x}
 \sum_{n\neq 0}\frac{1}
 {n^2\left(\x^2+\pi^2k^2n^2\right)^2}
 \right],\\
 \label{defq}
 Q=\frac{1}{\pi^2k^2}\sum_{n\neq 0}\frac{C_n^2-D_n^2}{n^2}
 &=&\frac{1}{2\pi^2k^2}
 \left[
 \frac{\x^4\cosh 2\x}{\sinh^2\x}\sum_{n\neq 0}\frac{\cos2\pi kn}
 {n^2\left(\x^2+\pi^2k^2n^2\right)^2}
 \right.\nonumber\\&&
 -\pi^2k^2\frac{\x^2\cosh 2\x}{\sinh^2\x}\sum_{n\neq 0}\frac{\cos2\pi kn}
 {\left(\x^2+\pi^2k^2n^2\right)^2}
 \nonumber\\&&
 +4\pi k\x^3\coth\x
 \sum_{n\neq 0}\frac{\sin2\pi kn}
 {n\left(\x^2+\pi^2k^2n^2\right)^2}
 \nonumber\\&&\left.
 +\frac{\x^2}{\sinh^2\x}
 \sum_{n\neq 0}\frac{\pi^2k^2n^2-\x^2}
 {n^2\left(\x^2+\pi^2k^2n^2\right)^2}
 \right].
 \end{eqnarray}
 The sums entering \rf{defp} and \rf{defq} can be calculated as
 $t\rightarrow 0$ or $t\rightarrow \infty$ limits of the following
 equality:
 \begin{eqnarray}\label{sum}
 &&\sum_{n\neq 0}\frac{\E^{2\pi ikn}}
 {(n+it)\left(\x^2+\pi^2k^2n^2\right)^2}
 =\frac{i}{\x^4t}
 -\frac{i\pi}{\x^4}
 \frac{\E^{(2s-1)\pi t}}
 {\left(1-\frac{\pi^2k^2t^2}{\x^2}\right)^2\sinh\pi t}
 \nonumber\\
 &&+\frac{i\pi }{2\x^4}\frac{2\sinh(2s-1)\frac{\x}{k}+\frac{\pi k t}{\x}
 \left(3-\frac{\pi^2k^2t^2}{\x^2}\right)
 \cosh(2s-1)\frac{\x}{k}}{\left(1-\frac{\pi^2k^2t^2}{\x^2}\right)^2\sinh
 \frac{\x}{k}}
 \nonumber\\
 &&-\frac{i\pi s}{k\x^3}\frac{\cosh(2s-1)\frac{\x}{k}+\frac{\pi k t}{\x}
 \sinh(2s-1)\frac{\x}{k}}{\left(1-\frac{\pi^2k^2t^2}{\x^2}\right)\sinh
 \frac{\x}{k}}
 +\frac{i\pi}{2k\x^3}\frac{\sinh 2s\frac{\x}{k}+\frac{\pi k t}{\x}
 \cosh 2s\frac{\x}{k}}{\left(1-\frac{\pi^2k^2t^2}{\x^2}\right)\sinh^2
 \frac{\x}{k}},
 \end{eqnarray}
 which can be obtained by Poisson resummation. Here $s$ is the
 fractional part of $k$. The result of the calculations reads:
 \begin{eqnarray}\label{mainp}
 P&=&\frac{1}{3k^2}\left(2s^3-3s^2+s\right)\coth\x
 -\frac{1}{6k}\left(6s^2-6s+1\right)\frac{\cosh 2 \x}{\x\sinh^2\x}
 \nonumber\\
 &&+\frac32(2s-1)\frac{\coth\x}{\x^2}
 -k\frac{\cosh 2\x}{\x^3\sinh^2\x}
 +\frac{1}{4k}\frac{\cosh 2\x}{\x\sinh^2\x}\frac{\cosh 2s\frac{\x}{k}}
 {\sinh^2\frac{\x}{k}}
 \nonumber\\
 &&-\frac{1}{2k}\frac{\coth\x}{\x}
 \frac{\sinh 2s\frac{\x}{k}}{\sinh^2\frac{\x}{k}}
 +\left(\frac{s}{k}\frac{\coth\x}{\x}+\frac34\frac{\cosh
 2\x}{\x^2\sinh^2\x}\right)
 \frac{\cosh(2s-1)\frac{\x}{k}}{\sinh\frac{\x}{k}}
 \nonumber\\
 &&-\left(\frac{s}{2k}\frac{\cosh 2\x}{\x\sinh^2\x}+\frac32\frac
 {\coth\x}{\x^2}\right)
 \frac{\sinh(2s-1)\frac{\x}{k}}{\sinh\frac{\x}{k}}
 \nonumber\\
 &&+\frac{1}{2\pi^2k}\frac{\x^3}{\sinh^2\x}
 \sum_{n\neq 0}\frac{1}
 {n^2\left(\x^2+\pi^2k^2n^2\right)^2},
 \\
 \label{mainq}
 Q&=&
 \frac{1}{6k^2}\left(6s^2-6s+1\right)\frac{\cosh 2 \x}{\sinh^2\x}
 -\frac{2}{k}(2s-1)\frac{\coth\x}{\x}+\frac32\frac{\cosh
 2\x}{\x^2\sinh^2\x} \nonumber\\
 &&-\frac{1}{2k^2}\frac{\cosh 2\x}{\sinh^2\x}\frac{\cosh 2s\frac{\x}{k}}
 {\sinh^2\frac{\x}{k}}+\frac{1}{k^2}\coth\x
 \frac{\sinh 2s\frac{\x}{k}}{\sinh^2\frac{\x}{k}}
 \nonumber\\
 &&-\left(\frac{2s}{k^2}\coth\x+\frac{1}{k}\frac{\cosh
 2\x}{\x\sinh^2\x}\right)
 \frac{\cosh(2s-1)\frac{\x}{k}}{\sinh\frac{\x}{k}}
 \nonumber\\
 &&+\left(\frac{s}{k^2}\frac{\cosh 2\x}{\sinh^2\x}+\frac{2}{k}\frac
 {\coth\x}{\x}\right)
 \frac{\sinh(2s-1)\frac{\x}{k}}{\sinh\frac{\x}{k}}
 \nonumber\\
 &&+\frac{\x^2}{2\sinh^2\x}
 \sum_{n\neq 0}\frac{\pi^2k^2n^2-\x^2}
 {\pi^2k^2n^2\left(\x^2+\pi^2k^2n^2\right)^2}.
 \end{eqnarray}

 The substitution of \rf{mainj02} into \rf{var} gives
 \begin{eqnarray}\label{gm2t}
 k\frac{\d\g_{I,t}^{(2)}}{\d k}&=&2k\left[
 \frac{\d}{\d k}\int_{0}^{\l k}d\x\,\x^2\left(\coth\x-
 \frac{\x}{\sinh^2\x}\right)P
 -\l^3k^2P(\l k)
 \right.\nonumber\\&&\left.
 -\int_{0}^{\l k}d\x\,\x^2\left(\coth\x-\frac{1}{\x}\right)
 \left(\coth\x-\frac{\x}{\sinh^2\x}\right)Q
 \right].
 \end{eqnarray}
 The leading infrared divergence comes from the first terms in $P$ and $Q$
 and is easily calculable:
 \begin{eqnarray}
 k\frac{\g_{I,t}^{(2)}}{\d k}&\sim&2k\left\{
 \frac{\d}{\d k}\left[\frac13\l^3k\frac13\left(2s^3-3s^2+s\right)\right]
 -\l^3\,\frac13\left(2s^3-3s^2+s\right)
 \right.\nonumber\\
 &&\left.-\frac23\l^3k\frac16\left(6s^2-6s+1\right)\right\}
 =-\frac49k\l^3\left(2s^3-3s^2+s\right).
 \nonumber
 \end{eqnarray}
 Integration over $k$ gives
 \begin{equation}\label{irdiv1}
 \ln\det\left(-D^2\right)\sim\frac29\l^3s^2(1-s)^2,
 \end{equation}
 the result, quoted in Sec. \ref{sec}.

 For integral $k$ \rf{irdiv1} turns to zero. One may expect that other
 power--like divergences also cancel. We shall see that this is really
 the case, moreover, the logarithmic divergence vanishes as well, and the
 resulting contribution to the determinant is finite and is of order
 $1/k$.

 After integration over $\x$ in \rf{gm2t} one should remove the variation
 with respect to $k$. This procedure is trivial for the first term and, as
 we are interested in the value of $\ln\det\left(-D^2\right)$ for integral
 $k$, we can put $s=0$ in the integrand. The second and the third terms
 should be integrated over $k$ directly. We find:
 \begin{eqnarray}\label{intp} &&\int_{0}^{\l
 k}d\x\,\x^2\left.\left(\coth\x-
 \frac{\x}{\sinh^2\x}\right)P\,\right|_{s=0} =-\frac16\l^2k-\frac32\left
 (\l
 k-\frac12\right) \nonumber\\ &&-2k\left(\ln\frac{\l
 k}{\pi}+\frac23+\gamma+2\z\right)
 +\frac12k\left(\ln\frac{k}{2\pi}+\frac53+\gamma+2\z\right)
 \nonumber\\
 &&+\frac32k\left(\ln\frac{k}{2\pi}+\l
 +\frac23+\gamma+2\z\right)
 -\frac32\left(-\l k+\frac12\right)
 +O\left(\frac{1}{k}\right)
 \\
 \label{intq}
 &&\int_{0}^{\l k}d\x\,\x^2\left(\coth\x-\frac{1}{\x}\right)
 \left(\coth\x-\frac{\x}{\sinh^2\x}\right)Q
 \nonumber\\
 &&
 =\frac{1}{6k^2}\left(6s^2-6s+1\right)
 \left(\frac23\l^3k^3-\l^2k^2\right)
 \nonumber\\
 &&
 -\frac{2}{k}(2s-1)\left(\frac12\l^2k^2-\l
 k+\frac13+\frac{\pi^2}{36}\right)
 \nonumber\\
 &&+3\left(\l k-\ln\frac{\l k}{\pi}-\frac{17}{12}-\gamma-2\z\right)
 \nonumber\\
 &&
 -\frac{1}{2k^2}\left[k^3\left(F_-^{\prime}(s)+\frac12sF_-^
 {\prime\prime}
 (s)\right)
 \right.\nonumber\\
 &&\left.+k^2\left(F_+(s)+sF_+^{\prime}(s)-2\ln\frac{k}{2\pi}
 -\frac{29}{6}-4\z\right)\right]
 \nonumber\\
 &&-\frac{1}{k^2}\left[\frac12k^3\left(F_+^{\prime}(s)
 +\frac12sF_+^{\prime\prime}(s)\right)+\frac12k^2\left(
 F_-(s)+sF_-^{\prime}(s)\right)-2sk\left(\frac13+\frac{\pi^2}
 {36}\right)\right]
 \nonumber\\
 &&-\frac{2s}{k^2}\left[k^3F_1^{\prime\prime}(\L)-k^2F_2^{\prime}(\L)
 +k\left(\frac13+\frac{\pi^2}{36}\right)\right]
 \nonumber\\
 &&-\frac{2}{k}\left[k^2F_2^{\prime}(\L)-k\left(
 \ln\frac{k}{2\pi}+F_1(\L)+\frac{17}{12}+2\z\right)\right]
 \nonumber\\
 &&+\frac{2s}{k^2}\sign(2s-1)\left(-k^3F_2^{\prime\prime}(\L)
 +k^2F_1^{\prime}(\L)\right)
 \nonumber\\
 &&+\frac{2}{k}\left[\sign(2s-1)\left(-k^2 F_1^{\prime}(\L)
 +kF_2(\L)\right)+(2s-1)\left(\frac13+\frac{\pi^2}{36}\right)\right]
 \nonumber\\
 &&
 +O\left(\frac{1}{k^2}\right),
 \end{eqnarray}
 where
 \begin{equation}\label{defdelta}
 \L(s)=1-|2s-1|,
 \end{equation}
 \begin{equation}\label{deff1}
 F_1(\L)=\frac{1-\E^{-\l\L}}{\L}-\frac12\left(
 \psi\left(1+\frac{\L}{2}\right)+\psi\left(1-\frac{\L}{2}\right)\right),
 \end{equation}
 \begin{equation}\label{deff2}
 F_2(\L)=\frac{\E^{-\l\L}}{\L}-\frac{\pi}{2}\cot\frac{\pi\L}{2},
 \end{equation}
 \begin{equation}\label{deffpm}
 F_{\pm}(s)=\psi(s)\pm\psi(-s).
 \end{equation}

 For the second term in \rf{gm2t} we have:
 \begin{eqnarray}\label{pinf}
 \l^3k^2P(\l k)&=&\frac13\left(2s^3-3s^2+s\right)\l^3
 -\frac13\left(6s^2-6s+1\right)\l^2
 +\frac32(2s-1)\l
 \nonumber\\
 &&-2+(1-\sign(2s-1))\left(\l^2s+\frac32\l\right)\E^{-\l\L},
 \end{eqnarray}
 but it gives zero contribution:
 \begin{equation}\label{intpinf}
 \int_0^1ds\,\l^3k^2P(\l k)=0.
 \end{equation}

 Let us consider the contribution of \rf{intq}. Some of the terms in this
 expression do not depend on $s$ explicitly and can be integrated over
 $k$ directly; others should be first integrated over $s$ from $0$ to $1$
 at fixed $N$ and then summed over $N$. Thus we obtain:
 \begin{eqnarray}\label{resq}
 &&\int^{k}dk'\,\int_0^{\l k'}d\x\,\x^2
 \left(\coth\x-\frac{1}{\x}\right)
 \left(\coth\x-\frac{\x}{\sinh^2\x}\right)Q
 \nonumber\\
 &&
 =\int^kdk'\,\left(3\l
 k'-3\ln 2\l+1-3\gamma+O\left(\frac{1}{k'^{2}}\right)\right)
 \nonumber\\
 &&+\sum_{N<k}\left(-\frac16\l^2-3\l N+\ln
 2\l-\frac12+3\gamma+O\left(\frac{1}{N^2}\right)\right)
 \nonumber\\&&
 =\left(-\frac16\l^2+\frac32\l-2\ln 2\l+\frac12\right)k+O\left(
 \frac{1}{k}\right)
 \end{eqnarray}
 The final result is given by a difference of \rf{intp} and \rf{resq} and
 is equal to zero up to the terms of order $1/k$.

 For the spatial components of the current we find
 \begin{equation}\label{coms}
 \left\langle v\left|\left[\d_{i}\hat{P},\E^{2\pi iknz}\right]
 \right|v\right\rangle
 =i\e_{ijk}n_j\s^k\frac{1}{\sinh\x}\left(C_n-\frac{\sin\pi nk}{\pi
 nk}\right)
 \end{equation}
 and
 \begin{equation}\label{js2}
 J_{i}^{(2)}=i\e_{ijk}n_j\s^k\frac{1}{4\pi^2}
 \frac{1}{\sinh\x}\sum_{n\neq 0}\left(\frac{C_n^2}{\pi^2k^2n^2}
 -\frac{C_n\sin\pi kn}{\pi^3k^3n^3}\right).
 \end{equation}
 The substitution of this expression in \rf{var} leads to the convergent
 integral, so the limit $k\rightarrow\infty$ and integration over
 $\x$ can be interchanged. Since $J_i^{(2)}$ is, at least, $o(1/k^3)$,
 it's contribution is negligible at large $k$. Thus we conclude that
 $\g_{I}^{(2)}=O(1/k)$.

\end{document}